\documentclass[preprint,12pt]{elsarticle}

\usepackage{amssymb}
\usepackage{amsmath}
\usepackage{array}
\usepackage{subfig}
\usepackage[pdftex]{hyperref}
\usepackage{subfloat}

\PassOptionsToPackage{unicode}{hyperref}
\PassOptionsToPackage{naturalnames}{hyperref}
\journal{Nuclear Physics B}

\begin{document}

\begin{frontmatter}



\title{Beam test result and digitization of TaichuPix-3: A Monolithic Active Pixel Sensors for CEPC vertex detector}

\author[a,c]{Hancen Lu}
\author[a,c]{Tianyuan Zhang}
\author[a,c]{Chang Xu}
\author[a,c]{Shuqi Li}
\author[a,c]{Xinhui Huang}
\author[a,c]{Ziyue Yan}
\author[a,c]{Jia Zhou}
\author[f]{Xiaoxu Zhang}
\author[f]{Yiming Hu}
\author[a,d]{Wei Wang}
\author[a,c]{Hao Zeng}
\author[a,c]{Xuewei Jia}
\author[a,d]{Zhijun Liang\corref{cor}}
\author[a,d]{Wei Wei\corref{cor}}
\author[a,d]{Ying Zhang\corref{cor}}
\author[b]{Tianya Wu}
\author[e]{Xiaomin Wei}
\author[f]{Ming Qi}
\author[f]{Lei Zhang}
\author[a,d]{Jun Hu}
\author[a,d]{Jinyu Fu}
\author[a,c,d]{Hongyu Zhang}
\author[a]{Gang Li}
\author[a]{Linghui Wu}
\author[a,c,d]{Mingyi Dong}
\author[a,d]{Xiaoting Li}
\author[h]{Raimon Casanova}
\author[g]{Liang Zhang}
\author[g]{Jianing Dong}
\author[e]{Jia Wang}
\author[e]{Ran Zheng}
\author[a,d]{Weiguo Lu}
\author[h,i]{Sebastian Grinstein}
\author[a]{Jo\~{a}o Guimar\~{a}es da Costa}


\cortext[cor]{Corresponding author: \href{mailto:liangzj@ihep.ac.cn}{liangzj@ihep.ac.cn} (Zhijun Liang), \href{mailto:zhangying83@ihep.ac.cn}{zhangying83@ihep.ac.cn} (Ying Zhang), \href{mailto:weiw@ihep.ac.cn}{weiw@ihep.ac.cn} (Wei Wei)}


\affiliation[a]{organization={Institute of High Energy Physics, Chinese Academy of Sciences},
            addressline={19B Yuquan Road, Shijingshan District}, 
            city={Beijing},
            postcode={100049},
            country={China}}
                      
\affiliation[b]{organization={School of Information Engineering, Nanchang University},
            city={Nanchang},
            postcode={330031},
            country={China}}

\affiliation[c]{organization={University of Chinese Academy of Sciences},
            addressline={19A Yuquan Road, Shijingshan District},
            city={Beijing},
            postcode={100049},
            country={China}}

\affiliation[d]{organization={State Key Laboratory of Particle Detection and Electronics},
            city={Beijing},
            postcode={100049},
            country={China}}

\affiliation[e]{organization={Northwestern Polytechnical University},
            city={Xi'an},
            country={China}}

\affiliation[f]{organization={Department of Physics, Nanjing University},
            city={Nanjing},
            postcode={210093},
            country={China}}

\affiliation[g]{organization={Institute of Frontier and Interdisciplinary Science and Key Laboratory of Particle Physics and Particle Irradiation, Shandong University},
            city={Qingdao},
            country={China}}

\affiliation[h]{organization={Institut  de Fisica d'Altes Energies (IFAE)},
            city={Bellaterra (Barcelona)},
            country={Spain}}

\affiliation[i]{organization={Catalan Institution for Research and Advanced Studies (ICREA)},
            city={Barcelona},
            country={Spain}}

\begin{abstract}
The Circular Electron-Positron Collider (CEPC), as the next-generation electron-positron collider, is tasked with advancing not only Higgs physics but also the discovery of new physics. Achieving these goals requires high-precision measurements of particles. Taichu seires, Monolithic Active Pixel Sensor (MAPS), a key component of the vertex detector for CEPC was designed to meet the CEPC's requirements. For the geometry of vertex detector is long barrel with no endcap, and current silicon lacks a complete digitization model, precise estimation of cluster size particularly causing by particle with large incident angle is needed. Testbeam results were conducted at the Beijing Synchrotron Radiation Facility (BSRF) to evaluate cluster size dependence on different incident angles and threshold settings. Experimental results confirmed that cluster size increases with incident angle. Simulations using the Allpix-Squared framework replicated experimental trends at small angles but exhibited discrepancies at large angles, suggesting limitations in linear electric field assumptions and sensor thickness approximations. The results from both testbeam and simulations have provided insights into the performance of the TaichuPix chip at large incident angles, offering a crucial foundation for the establishment of a digital model and addressing the estimation of cluster size in the forward region of the long barrel. Furthermore, it offers valuable references for future iterations of TaichuPix, the development of digital models, and the simulation and estimation of the vertex detector's performance.
\end{abstract}




\end{frontmatter}



\section{Introduction}

\begin{figure}[ht]
    \begin{center}
        \subfloat[]{\includegraphics[width=.5\linewidth]{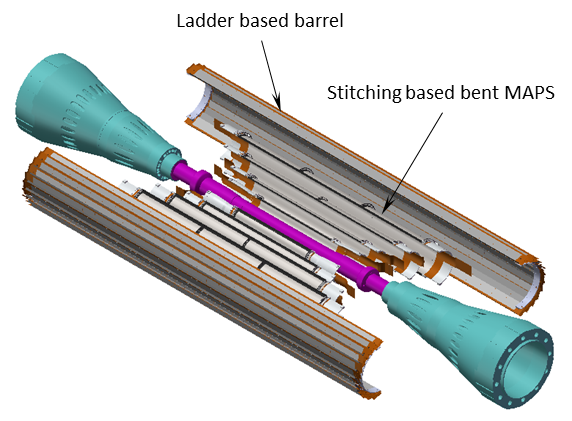}\label{fig:wholelayout} }
        \subfloat[]{\includegraphics[width=.5\linewidth]{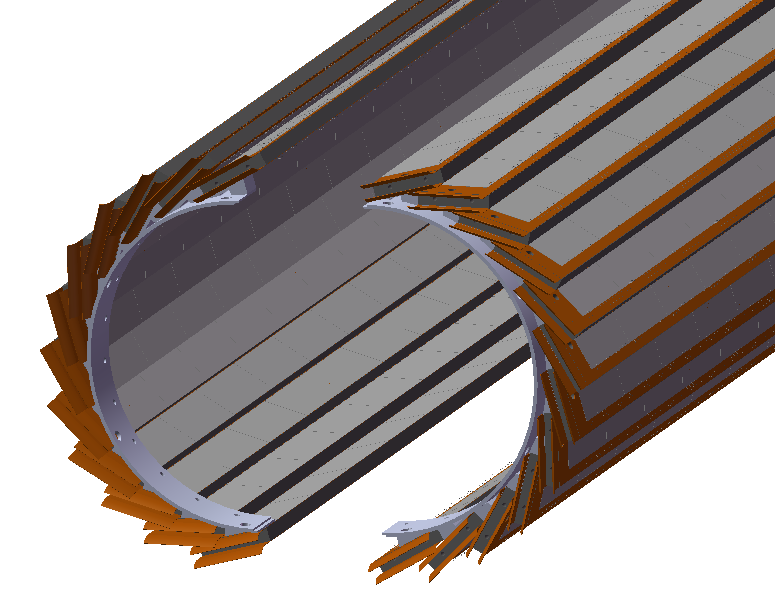}\label{fig:ladderlayout} }
        \caption{Baseline scheme of the CEPC vertex detector. (a) Structural design diagram of the vertex detector. The first four layers are designed as single-layer structures (full-model cylindrical structure) using 65 nm CIS stitched sensors, while the last two layers are designed as double-layer structure (a ladder structure) utilizing full size 65 nm CIS sensors. (b) Ladder based barrel layer assembly, where chip with no bent like TaichuPix-3 will be used on.}
        \label{fig:sitichinglayout}
    \end{center}
\end{figure}

The Circular Electron-Positron Collider (CEPC) focus on precisely measurement of Higgs boson, W boson and Z boson, as well as searching for new physics beyond Standard Model. Whether it is flavor physics or Higgs physics, precise vertex measurement is an essential requirement for achieving CEPC's physics goals. To meet the requirements of CEPC, the CDR \cite{CEPC} presents a serise of design specifications: single point resolution lower than $3~\mu \text{m}$, low power consumption below $50 ~\text{mW}\cdot\text{cm}^{-2}$ to enable forced air-cooling, low material budget below $0.15\% ~X/X_0$ per layer to minimize the effects of multiple scattering. Based on the requirements of CEPC for the vertex detector, whose geometry is shown as Figure~\ref{fig:sitichinglayout}, the CEPC vertex detector team has designed a data-driven, fast-readout Monolithic Active Pixel Sensor(MAPS), named TaichuPix. After several iterations(TaichuPix-1~\cite{taichupix-1}, TaichuPix-2~\cite{taichupix-2}), TaichuPix-3 has been developed. 

Since the vertex detector is one of the crucial sub-detector of CEPC detector, the performance of TaichuPix is quite important. As Figure~\ref{fig:sitichinglayout} shows, the structure of the CEPC vertex detector is a long barrel structure with no endcap, which means, the particle travels through the forward region will create a quite large cluster size. If the cluster size cannot be accurately estimated, it will significantly affect the background estimation and calculations of the vertex detector performance. This will lead to inaccurate estimates of these critical design metrics, which in turn impacts the design of the detector. However, current silicon-based chip in CEPCSW (a software developed for CEPC) lacks a complete digitization model,  resulting in low accuracy in simulations, particularly in forward region of the long barrel. Therefore, a beam test at the Beijing Synchrotron Radiation Facility (BSRF) was conducted to investigate the performance of TaichuPix at various incident angles, particularly at large angles. Based on the results of testbeam, a digital model of TaichuPix has been constructed using Allpix$^2$.

This article will discuss the results of the testbeam in BSRF and the Allpix$^2$ simulation, analyzing them to determine the relationship between the incident angle and cluster size, thereby laying the groundwork for the digitization of TaichuPix-3 in the future.

\section{Testbeam in BSRF}

In order to evaluate the performance of the TaichuPix-3 under different incident angles and threshold settings, beam test at BSRF was conducted. This section will briefly introduce the TaichuPix series of chips and then present our experimental setup and test results obtained from the BSRF test beam.

\subsection{TaichuPix}

Several prototype pixel sensors have been developed for the CEPC, such as the JadePix series~\cite{Chen_2019}\cite{DONG2023167967} with a rolling shutter readout architecture. In order to address the high hit density of the CEPC, a Monolithic Active Pixel Sensor (MAPS), named TaichuPix, has been developed with the goal of high readout speed and high spatial resolution. TaichuPix-1 and TaichuPix-2, are multi-project wafers, and TaichuPix-3 is a full-scale prototype with an engineering run. The pixel matrix of TaichuPix-3 is $1024\times 512$ with a pixel pitch of $25~\mu \text{m}$, and a thickness of $150~\mu \text{m}$. The proposed backup vertex detector consists of three layers of ladders, with double-sided mounted TaichuPix-3 sensors.

Each pixel of the TaichuPix-3 chip integrates a sensing diode, an analog front-end, and a digital logical readout in each pixel. The analog front-end is designed based on the ALPIDE~\cite{MAGER2016434} chip, which is developed for the upgrade of the ALICE Inner Tracking System (ITS)~\cite{LIU2024169355}\cite{REIDT2022166632}\cite{ITS2}\cite{ITS3}. In order to address the high hit rate of the CEPC, the analog front-end of TaichuPix-3 has been specifically optimized to ensure a quicker response. In addition, the digital logical readout includes a hit storage register, logic for pixel mask, and test pulse configuration. The digital logical readout follows the FE-I3~\cite{PERIC2006178} designed for the ATLAS pixel detector, but it has been modified to adjust the pixel address generator and relocate the timestamp storage from within the pixel to the end of the column. This modification was necessary due to pixel size constraints. Furthermore, the double-column drain peripheral readout architecture of the TaichuPix-3 chip employs an address encoder with a pull-up and pull-down network matrix.

This full-size prototype chip with techonology of CIS 180 nm, low dead time, low power density and low material (design specifications shown as Table~\ref{tab:Taichu Performance}) TaichuPix-3 has been tested in laboratory settings as well as in beam environments\cite{DESYIIBeamTest}.
\begin{table}[!htb]
	\centering
	\caption{TaichuPix-3 Performance Index.}
	\begin{tabular}{l l}
	  \hline
	  \textbf{Specification} & \textbf{Index}
	   \\
	  \hline
	    Pixel size & $25~\mu \text{m} \times 25~\mu \text{m} $ \\
	    Dimension & $15.9 ~\text{mm} \times 25.7 ~\text{mm}$  \\
	    Techonology & CIS $180 ~\text{nm}$ \\
	    Dead time & $< 500 ~\text{ns}$ \\
	    Power density & $< 200 ~\text{mW} \cdot \text{cm}^{-2}$ \\
	    Max. Hit rate & $36 \times 10^6 ~\text{cm}^{-2}\cdot\text{s}^{-1}$ \\
	  \hline
	\end{tabular}%
	\label{tab:Taichu Performance}
  \end{table}%

\subsection{Cluster size}
\label{subsec:clustersize}

If the chip is an ideal two-dimensional plane, when a particle passes through it, the corresponding pixel generates a signal, and the pixel position is the exact location of a hit caused by the particle. However, in practical situations, due to the small size of TaichuPix-3's pixels and the finite thickness of the chip, the particle's trajectory within the chip can simultaneously hit multiple pixels, each producing a signal. Additionally, when a particle passes through the chip, it generates electron-hole pairs that drift toward the collection electrode under the influence of an electric field, resulting in signals. Various physical effects during this process (such as charge sharing effects) can also cause neighboring pixels near the hit position to generate signals. Due to the influence of these physical effects, it is difficult to directly determine which pixel corresponds to the actual path taken by the particle. If this is not addressed, it can lead to degraded spatial resolution. For chips like TaichuPix, which do not record signal amplitudes, it is generally assumed that the center point of the trajectory corresponds to its geometric center.

\begin{figure}[!htb]
    \centering
    \includegraphics[width = .7\textwidth]{ 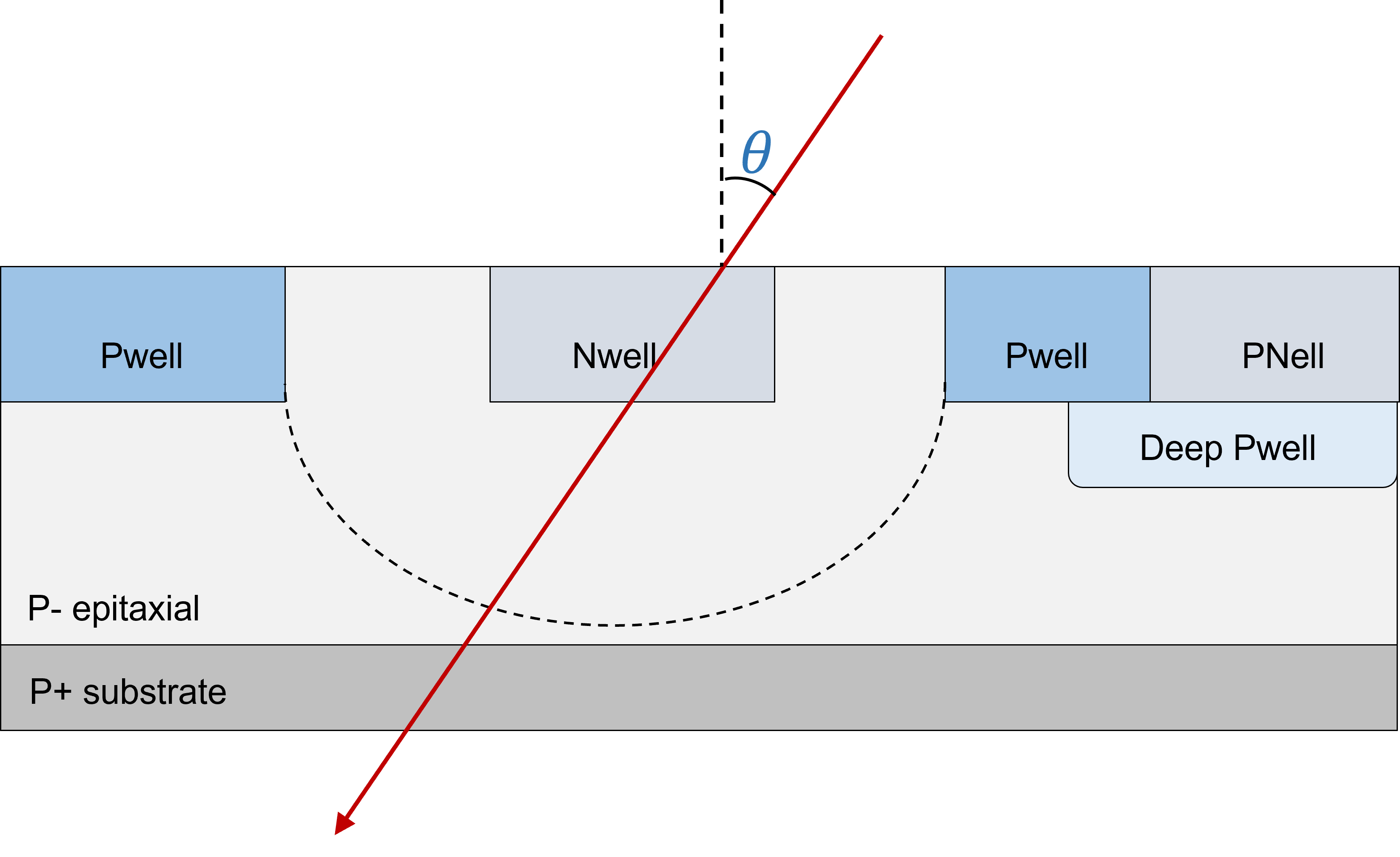}
    \caption{Particle track through chip. The incident angle $\theta$ is defined as the }
    \label{fig:trackDraw}
\end{figure}

The factors affecting the cluster size could be roughly categorize into two types: first, multiple pixels are hit during the particle's motion within the chip, leading to the generation of many signals; second, physical effects such as charge sharing cause neighboring pixels to generate signals. For the latter, thresholds could be set during readout to mitigate the impact. Therefore, it can be considered that the primary influencing factor of the cluster size in the beam test is the particle's trajectory. In this beam test, there is no external magnetic field, so the particle's motion can be approximated as a straight line. Additionally, the chip is a flat cube, so the particle's trajectory is only related to the incident angle. In other words, the only thing should be considered is the effect of the incident angle $\theta$ on the cluster size. To this end, a simple model is established to describe the relationship between cluster size and incident angle $\theta$:

\begin{equation}
    C = a \sec \theta + b
    \label{eq:fittingmodel}
\end{equation}
where $C$ stand for cluster size and $\theta$ means the incident angle, as Fig.~\ref{fig:trackDraw} shows.

\subsection{Test beam setup}

\begin{figure}[!htb]
    \centering
    \includegraphics[width = .7\textwidth]{ 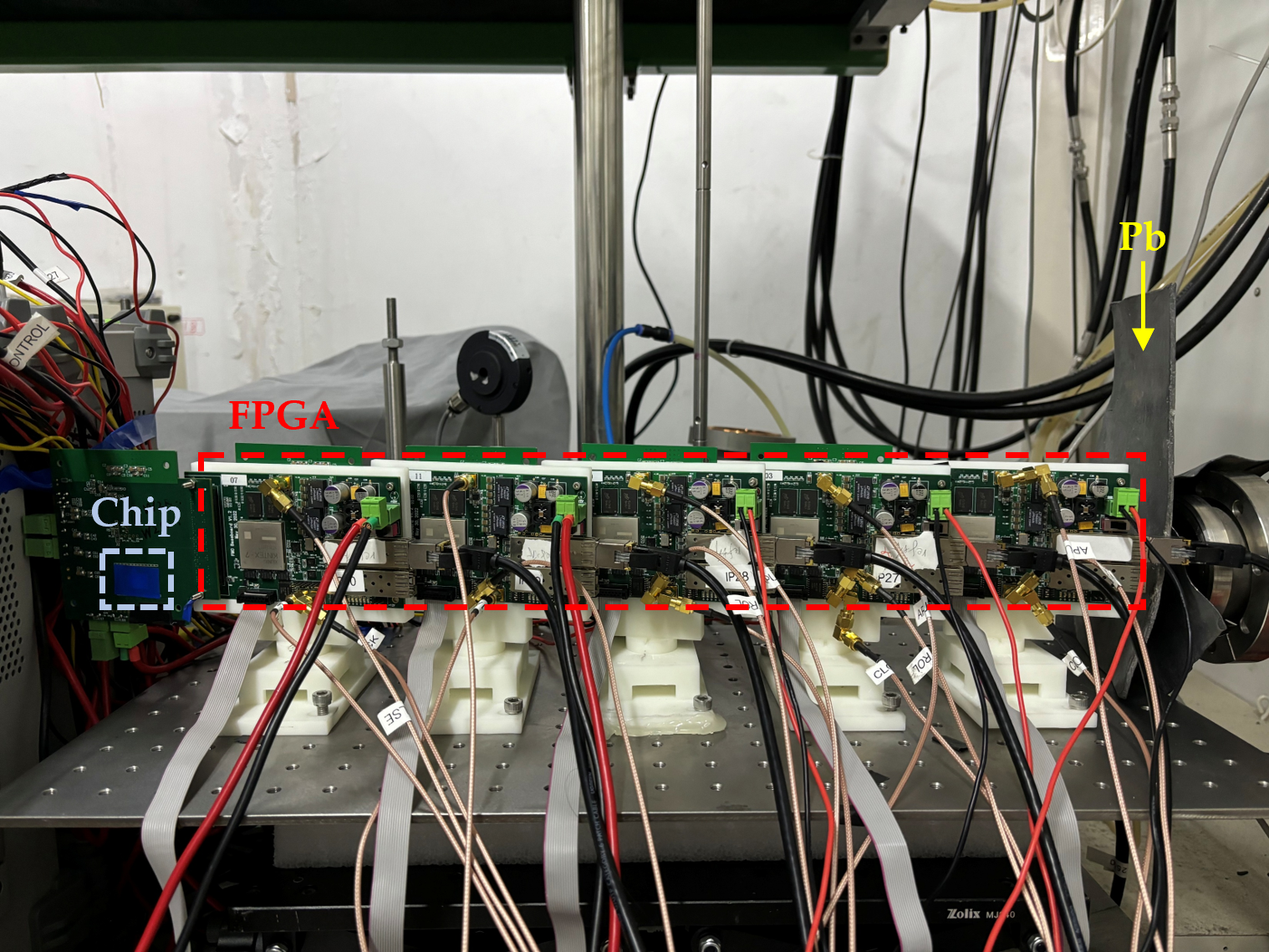}
    \caption{BSRF beam experimental setup.}
    \label{fig:setup}
\end{figure}

The experiment was conducted at the 4W2 high-pressure test station of the BSRF to assess beam tests. X-rays emitted were shielded with a specific thickness of lead. The energy of the emitted electrons was 2.5 GeV. As shown in Fig.~\ref{fig:setup}, the experimental setup involved placing chips at fixed angles relative to the electrons, arranged outward at 25 mm intervals from the particle emission point. These chips were sequentially connected to an interposer board, an FPGA readout board, and a SiTCP protocol Ethernet port, forming a comprehensive readout system. Data packets were transmitted through the Ethernet port to a switch and then sent to the host computer for further analysis, with data collection facilitated by a dedicated DAQ system.

\begin{figure}[!htb]
    \centering
    \includegraphics[width = .7\textwidth]{ 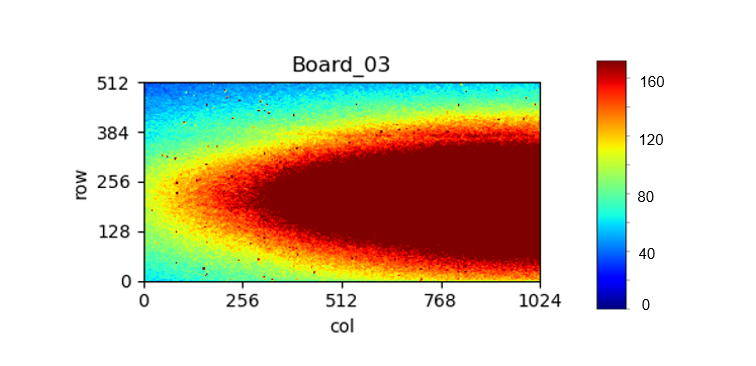}
    \caption{Hitmap of board 3 with 2.5 GeV electrons at $\theta = 60^\circ$.}
    \label{fig:exphitmap}
\end{figure}

During the beam test, the TaichuPix-3 sensor operated in triggerless mode with a back bias set at 0 V. The readout system functioned reliably throughout all runs, achieving a maximum data rate of approximately 6 MB/s. Figure~\ref{fig:exphitmap} illustrates an example of a hit diagram, demonstrating the normal operation of the overall inspection system. The incident angles are set at $0^\circ$, $60^\circ$,  $70^\circ$ and $80^\circ$. After completing a data acquisition experiment at each fixed angle, the five chips are uniformly adjusted to align with the angle of electron emission. Each experiment at a fixed angle is conducted separately at ITHR=32 and 64.

The offline analysis software used in this article is same as the DESY II beam test~\cite{DESYIIBeamTest}. The software primarily includes decoding raw data, clustering alignment, as well as track fitting. Due to the presence of too many background photons, alignment becomes difficult to achieve, making it challenging to ensure the accuracy of the results that rely on alignment. Therefore, in this section, cluster size under different conditions will be discussed.

\subsection{Beam test result}

In an ideal situation, when a particle passes through the chip, it creates electron-hole pairs, which drift under the influence of an electric field and ultimately form a signal at the collecting electrode. However, in reality, signals can still be generated within the chip for various reasons even in the absence of particles. Therefore, a threshold needed to be set to suppress or even eliminate these noise signals. The magnitude of the signal itself mainly depends on the energy loss of the particle as it traverses the chip. If the energy loss is relatively small at a certain pixel (for example, when the trajectory is very short), the number of electron-hole pairs generated may be overwhelmed by the noise signals, leading to the signal being truncated along with the noise. This means that if the threshold is set too high, a portion of the signal may be lost, thereby affecting the performance of the detector. Conversely, if the threshold is set too low, too much noise may be mistakenly recorded as signal, which would also impact the detector's performance. Thus, the appropriate threshold setting is needed to be carefully considered. In this beam test, ITHR 32, 64, 96 respectively correspond to threshold  $368~e^-$, $432~e^-$, $491~e^-$.

\begin{figure}[!htb]
    \centering
    \includegraphics[width = .7\textwidth]{ 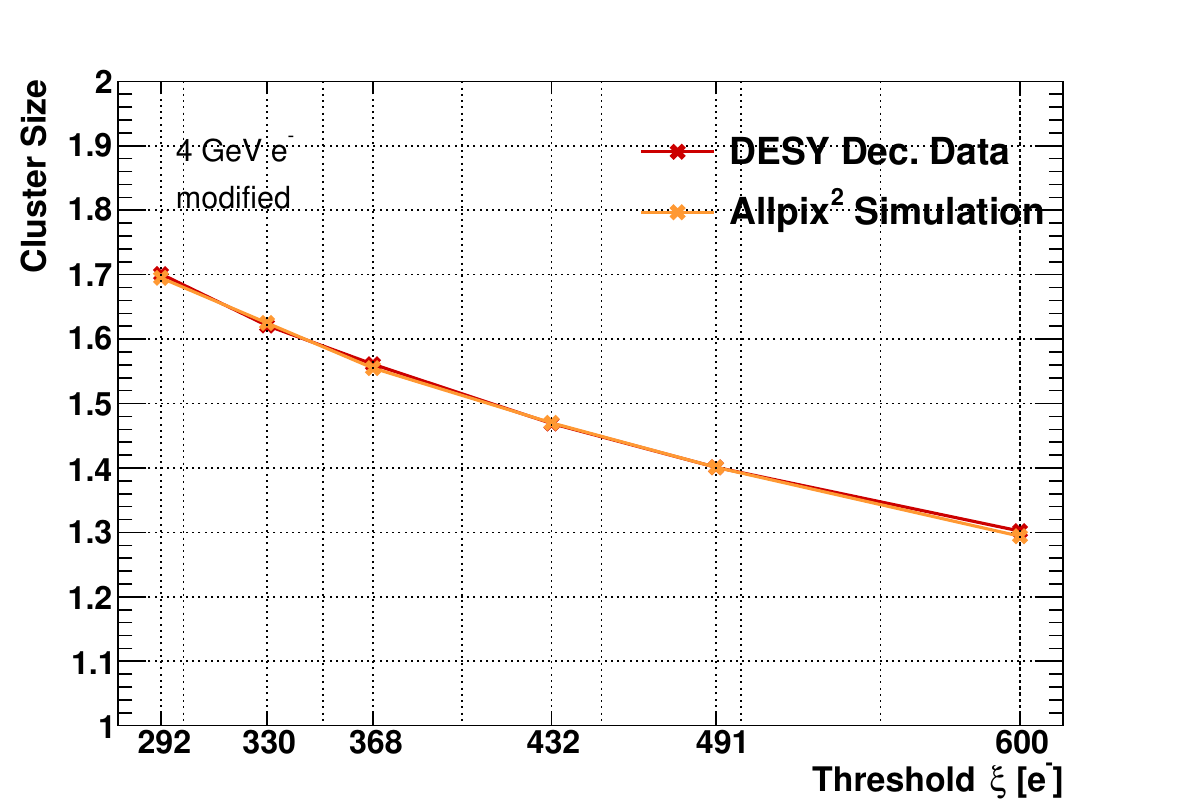}
    \caption{Cluster size vs. Threshold @ DESY II testbeam result~\cite{DESYIIBeamTest}.}
    \label{fig:DESYAllPix}
\end{figure}

As seen from the Fig.~\ref{fig:DESYAllPix}, when the threshold increases, the cluster size decrease, which is consistent with our hypothesis. Considering that overly aggressive threshold settings could result in excessive signal loss, the case of ITHR 96 would not be taken into account in our subsequent studies on the impact of incident angle on cluster size.

\begin{figure}[!htb]
    \centering
    \includegraphics[width = .7\textwidth]{ 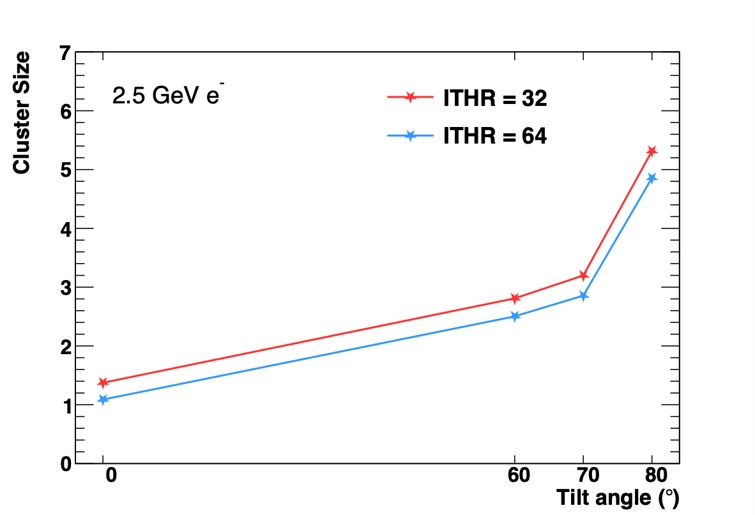}
    \caption{BSRF beam test result.}
    \label{fig:beamtestResult}
\end{figure}

The correlation between the incident angle and cluster size was mentioned in the section~\ref{subsec:clustersize}. In this beam test, there are four different incident angles: $0^\circ$, $60^\circ$, $70^\circ$, and $80^\circ$ to investigate the quantitative relationship between incident angle and cluster size. As Fig.~\ref{fig:beamtestResult} shows, with the increase in the incident angle, the cluster size also increases, which is consistent with our predictions stated in the section~\ref{subsec:clustersize}.

\section{Simulation based on Allpix-Squared}

To establish a reliable digital model for TaichuPix-3 and the future Taichu series of chips, rapid simulations are essential. Among various simulation software, Allpix$^2$ is chosen to be used for the simulation of TaichuPix-3. This section will present the simulation of TaichuPix-3 using Allpix$^2$, including the simulation setup, the simulation results, and a comparison between the simulation results and the test beam results.

\subsection{Simulation setup in Allpix-Squared}

Allpix$^2$ ~\cite{AP2} is a generic, open-source software framework for the simulation of silicon pixel detectors. Its goal is to ease the implementation of detailed simulations for both single detectors and more complex setups such as beam telescopes from incident radiation to the digitised detector response.

For some reasons, it is difficult to completely remap the simulation model for TaichuPix-3. Therefore, the preset model was used during the simulation process. Aside from some unspecified parameters, the main parameters of our model and simulation are as Table~\ref{tab:Taichupix-3 model} . In order to simplify the simulation process, the construction of the detector geometry consists of six layers of TaichuPix-3, with a spacing of 40 mm between each layer and the tilt angles being $0^\circ$, $16^\circ$, $60^\circ$, $65^\circ$, $70^\circ$ and $80^\circ$.

\begin{table}[!htb]
	\centering
	\caption{Main parameters of TaichuPix-3 and simulation.}
	\begin{tabular}{l l}
	  \hline
	  \textbf{Name} & \textbf{Index}
	   \\
	  \hline
	    Type & monolithic \\
	    Geometry & pixel \\
	    Pixel size & $25~\mu \text{m} \times 25~\mu \text{m} $ \\
	    Number of pixels & $1024 \times 512$ \\
	    Sensor thickness & -- \\
	    Chip thickness & $150 ~\mu \text{m}$ \\
	  \hline
	   Particle type & $e^-$\\
	   Source energy & 2.5 GeV \\
	   Source type & beam \\
	   Physics list & FTFP\_ BERT\_ LIV \\
	  \hline
	    Electric field model & linear\\
	    Electric field bias voltage & -- \\
	    Integration time & $15 ~\text{ns}$\\
	    Electronics noise & $20~e^-$ \\
	    Threshold & -- \\
	  \hline
	\end{tabular}%
	\label{tab:Taichupix-3 model}
  \end{table}

\subsection{Simulation result}

In the design of detectors, physical simulation is a very important task. A complete physical simulation can effectively assess the performance of the detector and provide references for improvement directions. To achieve this, the CEPC team has developed the CEPCSW software framework 
. The resources consumed by a complete full detector simulation are enormous, which necessitates the digitization of chips to simulate the detector performance in real situations. The results presented in this section are based on the beam test simulation results of the chip digitization using Allpix$^2$.


\begin{figure}[!htb]
    \centering
    \includegraphics[width = .7\textwidth]{ 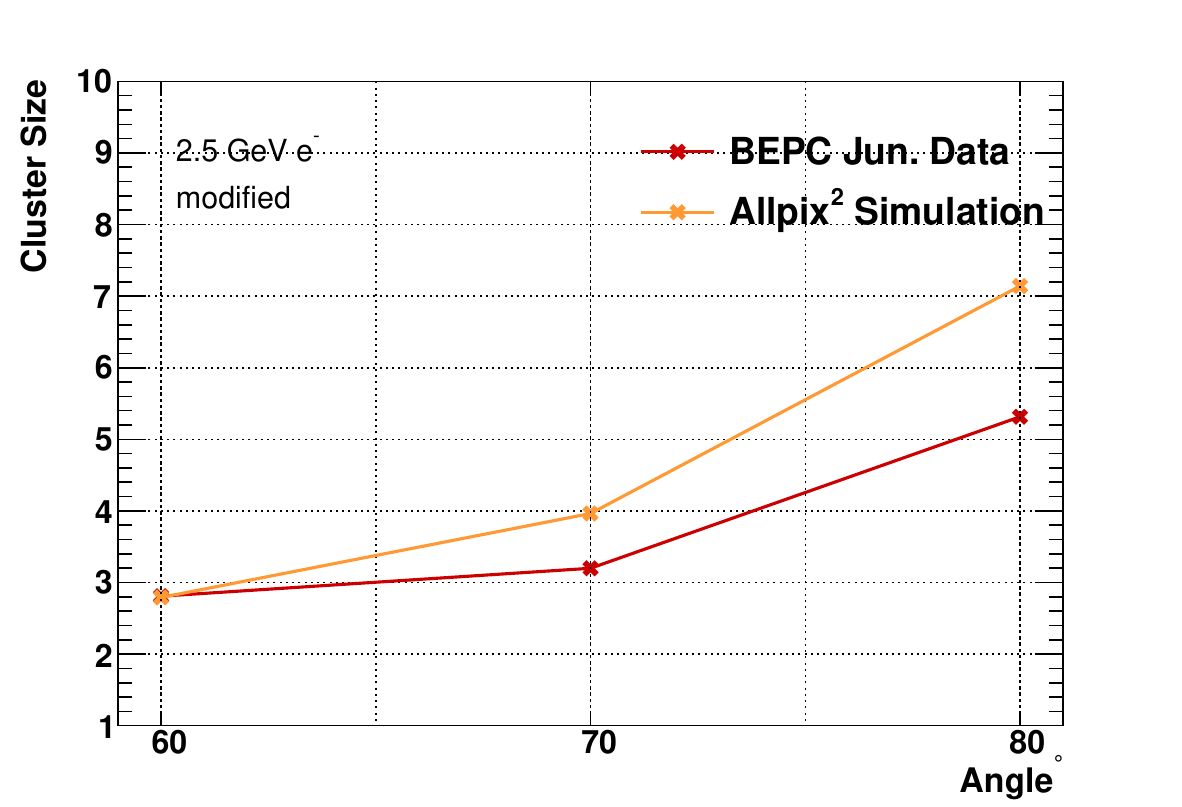}
    \caption{Cluster size vs. Tilt Angle @ BSRF testbeam result.}
    \label{fig:BSRFAllPix}
\end{figure}

As shown in Fig.~\ref{fig:DESYAllPix}, under the consistent conditions of other test parameters, the relationship between Cluster size and Threshold has a good agreement between the simulation results and experimental results. This indicates that if the conditions are set appropriately, Allpix$^2$ can provide results that match the experiments. For the relationship between Cluster size and Angle, the results obtained using parameters identical to those of the experiment are shown in Fig.~\ref{fig:BSRFAllPix}. Although both sets of results align well at the starting point, the gap between the experimental and simulated cluster sizes gradually increases as the Angle increases.

\subsection{Results analysis}

\begin{figure}[!tpb]
    \centering
    \subfloat[ITHR 32]
            {\label{fig:fitres1}
    		\includegraphics[width=0.5\textwidth]{ 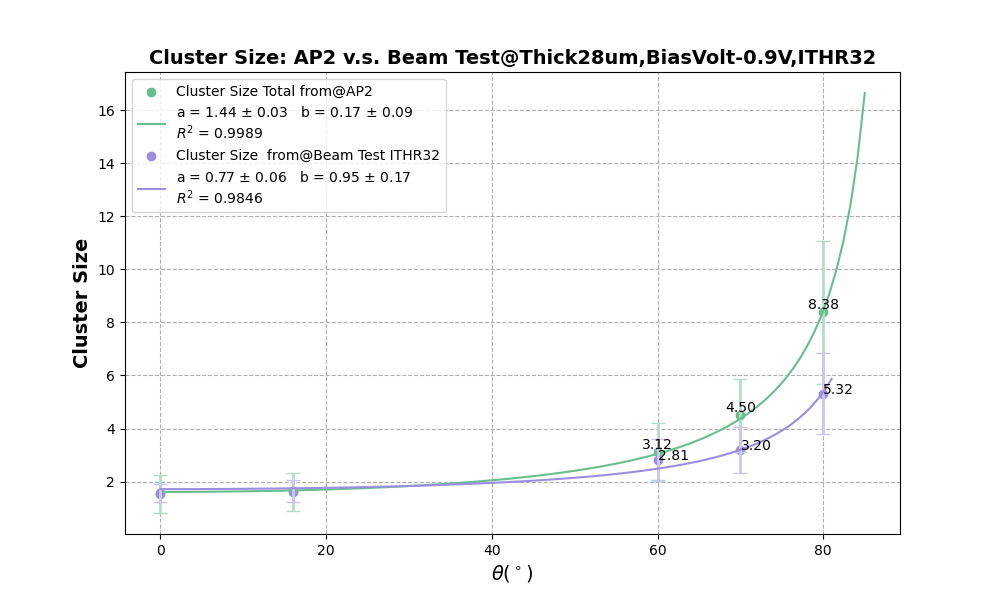}}
    \subfloat[ITHR 64]
            {\label{fig:fitres2}
    		\includegraphics[width=0.5\textwidth]{ 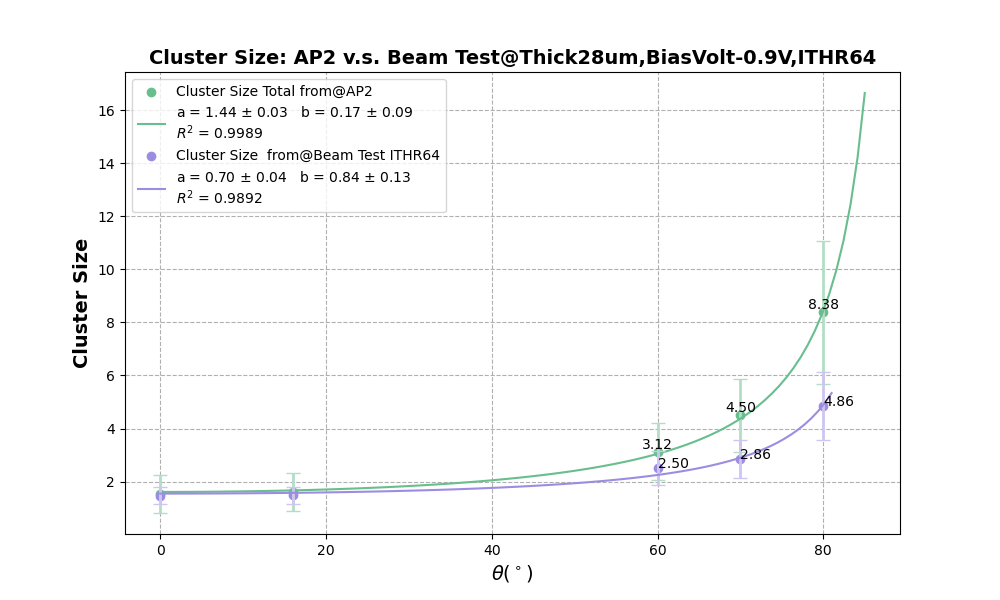}}
    \caption{Fitting result using model Eq.~\ref{eq:fittingmodel}, with sensor thickness $28 \mu\text{m}$ and bias voltage $-0.9$V}
    \label{fig:fitres}
\end{figure}

In the previous VXD prototype beam test~\cite{DESYIIBeamTest} at DESYII, the same TaichuPix-3 chip was used, and the testing environment was similar, so the angle data can be used for this analysis. 
Using the fitting model as Eq.~\ref{eq:fittingmodel}, the result show as Fig.~\ref{fig:fitres}.

For different threshold parameters, the fitting results of the Allpix$^2$ simulation are quite good. In the case of the beam test data, although there are some deviations around $\theta = 60^\circ$ that affect the fitting to some extent, it can still be concluded that the primary factor influencing the cluster size for different theta values is the number of pixels traversed by the track. Comparing the simulation and experimental results, similar to those in Fig~\ref{fig:DESYAllPix}, this set of simulation parameters matches the experimental results well at small angles. However, as the angle increases, the discrepancy between the simulation and the beam test experimental data gradually increases. At large angles ($>70^\circ$), the mean difference in Cluster size reaches 3.04. This is partly due to the significant broadening of the Cluster size distribution at larger angles and also indicates that the settings of the simulation parameters may differ from those in the actual experiment.

\begin{figure}[!htb]
    \centering
    \includegraphics[width = \textwidth]{ 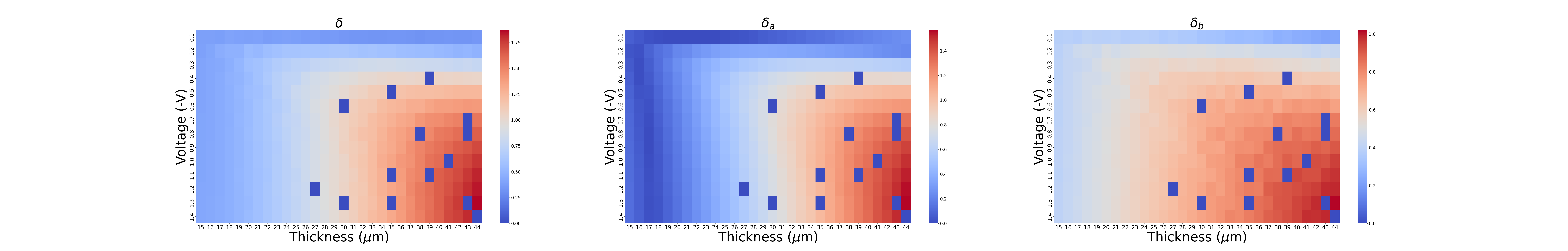}
    \caption{Difference between simulation and experiment with different simulation parameters. }
    \label{fig:gridsimerror}
\end{figure}

One possible reason is that, due to manufacturing processes, the approximation of a linear electric field during the drift of electron-hole pairs is not ideal. However, since the chips within the same module of the entire detector are of the same specifications, finding a suitable set of parameters as an overall approximation is reasonable. Another possible reason is that the substrate also contributes a portion of the electron-hole pairs, so the actual sensitive thickness may differ from the given parameters. Considering these two points, thickness and bias voltage were treated as variables and performed grid simulations within the voltage range of $-0.1 ~\text{V} \to -1.5 ~\text{V}$ and thickness range of $15 ~\mu \text{m} \to 45 ~\mu \text{m} $ to obtain results, using $\delta = \sqrt{\delta_a^2 + \delta_b^2 }$(where $\delta_a = a_{\text{sim}}-a_{\text{exp}}$ and $\delta_b = b_{\text{sim}}-b_{\text{exp}}$) as a measure of the deviation from the experimental degree, as shown in the Fig.~\ref{fig:gridsimerror}

\begin{figure}[!htb]
    \centering
    \includegraphics[width = .7\textwidth]{ 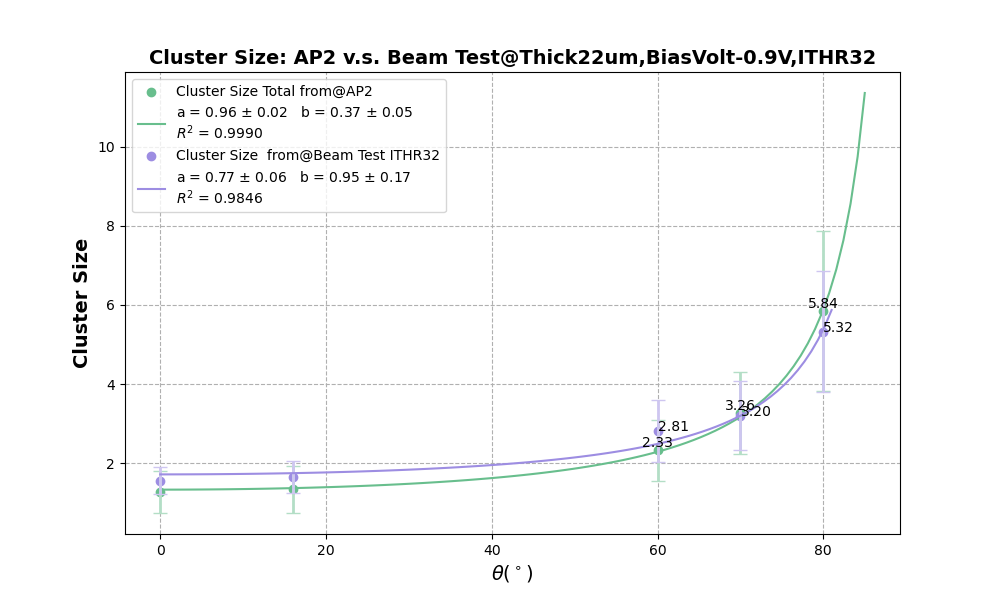}
    \caption{Simulation result with sensor thickness 23$\mu$m and bias voltage -0.9V}
    \label{fig:simulationCorrect}
\end{figure}

As mentioned earlier, due to some discrepancies between our model and the actual chip, it is unlikely that the simulation results will match the experiments exactly. In fact, considering that the main errors in our estimation of cluster size occur at large incident angles, the simulation results would not be required to perfectly align with the experiments; it is sufficient for the simulation to provide approximate estimates of cluster size under large incident angles. Taking these factors into account, a set of parameters that meet our requirements were filtered out , as shown in Figure~\ref{fig:simulationCorrect}, by referring to Figure~\ref{fig:gridsimerror} and excluding some parameters that do not align with actual conditions (such as low voltage or excessive thickness). It can be seen that although there is some deviation from the experimental results at small incident angles, the simulation agrees well with the actual results when the incident angle is greater than 60$^\circ$.

\section{Conclusions}


To better explore the performance of the series of chips developed for the CEPC vertex detector, particularly TaichuPix-3 under large incident angles, angle-dependent beam tests were conducted using a 2.5 GeV electron beam at BSRF. Based on the experimental results, a digital model of cluster size as a function of incident angle was established . At incident angles of 0$^\circ$, 60$^\circ$, 70$^\circ$, and 80$^\circ$ with an ITHR of 32, the corresponding cluster sizes were 1.56, 2.81, 3.20, and 5.31, respectively. Additionally, simulations were performed using Allpix$^2$. Due to the uncertainty of the doping parameters, two variable parameters were scanned : thickness of sensitive area and bias voltage. By considering both experimental and simulation data, the reasonableness of the model was validated . Through the analysis of the simulation results from the parameter scan, an Allpix$^2$ model aligning well with the experimental results at large incident angles was obtained . Based on these findings, a complete digital algorithm for the Taichu series chip is planned to be developed in the future and applied to precisely calculate beam background and physical performance in the CEPC vertex detector.


\begin{thebibliography}{00}



\bibitem{CEPC}Group, T. CEPC Conceptual Design Report: Volume 2 - Physics \& Detector.  (2018), https://arxiv.org/abs/1811.10545
\bibitem{DESYIIBeamTest}Wu, T., Li, S., Wang, W., Zhou, J., Yan, Z., Hu, Y., Zhang, X., Liang, Z., Wei, W., Zhang, Y., Wei, X., Huang, X., Zhang, L., Qi, M., Zeng, H., Jia, X., Hu, J., Fu, J., Zhang, H., Li, G., Wu, L., Dong, M., Li, X., Casanova, R., Zhang, L., Dong, J., Wang, J., Zheng, R., Lu, W., Grinstein, S. \& Da Costa, J. Beam test of a 180nm CMOS Pixel Sensor for the CEPC vertex detector. {\em Nuclear Instruments And Methods In Physics Research Section A: Accelerators, Spectrometers, Detectors And Associated Equipment}. \textbf{1059} pp. 168945 (2024), https://www.sciencedirect.com/science/article/pii/S0168900223009452
\bibitem{taichupix-2}Wei, X., Wei, W., Wu, T., Zhang, Y., Li, X., Zhang, L., Lu, W., Liang, Z., Dong, J., Li, L., Wang, J., Zheng, R., Casanova, R., Grinstein, S., Hu, Y. \& Costa, J. High data-rate readout logic design of a 512 × 1024 pixel array dedicated for CEPC vertex detector. {\em Journal Of Instrumentation}. \textbf{14}, C12012 (2019,12), https://dx.doi.org/10.1088/1748-0221/14/12/C12012
\bibitem{taichupix-1}Wu, T., Wei, W., Grinstein, S., Casanova, R., Zhang, Y., Wei, X., Dong, J., Zhang, L., Li, X., Liang, Z., Costa, J., Lu, W., Li, L., Wang, J., Zheng, R., Yang, P. \& Huang, G. The TaichuPix1: a monolithic active pixel sensor with fast in-pixel readout electronics for the CEPC vertex detector. {\em Journal Of Instrumentation}. \textbf{16}, P09020 (2021,9), https://dx.doi.org/10.1088/1748-0221/16/09/P09020
\bibitem{AP2}Spannagel, S., Wolters, K., Hynds, D., Alipour Tehrani, N., Benoit, M., Dannheim, D., Gauvin, N., Nürnberg, A., Schütze, P. \& Vicente, M. Allpix2: A modular simulation framework for silicon detectors. {\em Nuclear Instruments And Methods In Physics Research Section A: Accelerators, Spectrometers, Detectors And Associated Equipment}. \textbf{901} pp. 164-172 (2018), https://www.sciencedirect.com/science/article/pii/S0168900218307411
\bibitem{MAGER2016434}Mager, M. ALPIDE, the Monolithic Active Pixel Sensor for the ALICE ITS upgrade. {\em Nuclear Instruments And Methods In Physics Research Section A: Accelerators, Spectrometers, Detectors And Associated Equipment}. \textbf{824} pp. 434-438 (2016), https://www.sciencedirect.com/science/article/pii/S0168900215011122, Frontier Detectors for Frontier Physics: Proceedings of the 13th Pisa Meeting on Advanced Detectors
\bibitem{LIU2024169355}Liu, J. ALICE ITS3: A truly cylindrical vertex detector based on bent, wafer-scale stitched CMOS sensors. {\em Nuclear Instruments And Methods In Physics Research Section A: Accelerators, Spectrometers, Detectors And Associated Equipment}. \textbf{1064} pp. 169355 (2024), https://www.sciencedirect.com/science/article/pii/S016890022400281X
\bibitem{REIDT2022166632}Reidt, F. Upgrade of the ALICE ITS detector. {\em Nuclear Instruments And Methods In Physics Research Section A: Accelerators, Spectrometers, Detectors And Associated Equipment}. \textbf{1032} pp. 166632 (2022), https://www.sciencedirect.com/science/article/pii/S0168900222002042
\bibitem{PERIC2006178}Perić, I., Blanquart, L., Comes, G., Denes, P., Einsweiler, K., Fischer, P., Mandelli, E. \& Meddeler, G. The FEI3 readout chip for the ATLAS pixel detector. {\em Nuclear Instruments And Methods In Physics Research Section A: Accelerators, Spectrometers, Detectors And Associated Equipment}. \textbf{565}, 178-187 (2006), https://www.sciencedirect.com/science/article/pii/S0168900206007649, Proceedings of the International Workshop on Semiconductor Pixel Detectors for Particles and Imaging
\bibitem{Chen_2019}Chen, L., Zhang, Y., Zhu, H., Ai, X., Fu, M., Kiuchi, R., Liu, Y., Liu, Z., Lou, X., Lu, Y., Ouyang, Q., Shi, X., Tao, J., Wang, K., Wang, N., Yang, C. \& Zhou, Y. Characterization of the prototype CMOS pixel sensor JadePix-1 for the CEPC vertex detector. {\em Journal Of Instrumentation}. \textbf{14}, C05023 (2019,5), https://dx.doi.org/10.1088/1748-0221/14/05/C05023
\bibitem{DONG2023167967}Dong, S., Yang, P., Zhang, Y., Zhou, Y., Wang, H., Xiao, L., Zhang, L., Shi, Z., Guo, D., Wu, Z., Dong, J., Lu, Y., Sun, X. \& Ouyang, Q. Design and characterisation of the JadePix-3 CMOS pixel sensor. {\em Nuclear Instruments And Methods In Physics Research Section A: Accelerators, Spectrometers, Detectors And Associated Equipment}. \textbf{1048} pp. 167967 (2023), https://www.sciencedirect.com/science/article/pii/S0168900222012591
\bibitem{ITS2}Abelev, B., Adam, J., Adamová, D., Aggarwal, M., Aglieri Rinella, G., Agnello, M., Agostinelli, A., Agrawal, N., Ahammed, Z., Ahmad, N., Ahmad Masoodi, A., Ahmed, I., Ahn, S., Ahn, S., Aimo, I., Aiola, S., Ajaz, M., Akindinov, A., Aleksandrov, D., Alessandro, B., Alexandre, D., Alici, A., Alkin, A., Alme, J., Alt, T., Altini, V., Altinpinar, S., Altsybeev, I., Alves Garcia Prado, C., Anderssen, E., Andrei, C., Andronic, A., Anguelov, V., Anielski, J., Anticic, T., Antinori, F., Antonioli, P., Aphecetche, L., Appelshäuser, H., Arbor, N., Arcelli, S., Armesto, N., Arnaldi, R., Aronsson, T., Arsene, I., Arslandok, M., Augustinus, A., Averbeck, R., Awes, T., Azmi, M., Bach, M., Badalà, A., Baek, Y., Bagnasco, S., Bailhache, R., Bairathi, V., Bala, R., Baldisseri, A., Baltasar Dos Santos Pedrosa, F., Bán, J., Baral, R., Barbera, R., Barile, F., Barnaföldi, G., Barnby, L., Barret, V., Bartke, J., Basile, M., Bastian Van Beelen, J., Bastid, N., Basu, S., Bathen, B., Batigne, G., Battistin, M., Batyunya, B., Batzing, P., Baudot, J., Baumann, C., Bearden, I., Beck, H., Bedda, C., Behera, N., Belikov, I., Bellini, F., Bellwied, R., Belmont-Moreno, E., Bencedi, G., Benettoni, M., Benotto, F., Beole, S., Berceanu, I., Bercuci, A., Berdnikov, Y., Berenyi, D., Berger, M., Bertens, R., Berzano, D., Besson, A., Betev, L., Bhasin, A., Bhati, A., Bhatti, A., Bhattacharjee, B., Bhom, J., Bianchi, L., Bianchi, N., Bianchin, C., Bielcík, J., Bielcíková, J., Bilandzic, A., Bjelogrlic, S., Blanco, F., Blau, D., Blume, C., Bock, F., Boehmer, F., Bogdanov, A., Boggild, H., Bogolyubsky, M., Boldizsár, L., Bombara, M., Book, J., Borel, H., Borissov, A., Bornschein, J., Borshchov, V., Bortolin, C., Bossú, F., Botje, M., Botta, E., Böttger, S., Braun-Munzinger, P., Breitner, T., Broker, T., Browning, T., Broz, M., Bruna, E., Bruno, G., Budnikov, D., Buesching, H., Bufalino, S., Buncic, P., Busch, O., Buthelezi, Z., Caffarri, D., Cai, X., Caines, H., Caliva, A., Calvo Villar, E., Camerini, P., Canoa Roman, V., Carena, F., Carena, W., Cariola, P., Carminati, F., Casanova Díaz, A., Castillo Castellanos, J., Casula, E., Catanescu, V., Caudron, T., Cavicchioli, C., Ceballos Sanchez, C., Cepila, J., Cerello, P., Chang, B., Chapeland, S., Charvet, J., Chattopadhyay, S., Chattopadhyay, S., Cherney, M., Cheshkov, C., Cheynis, B., Chibante Barroso, V., Chinellato, D., Chochula, P., Chojnacki, M., Choudhury, S., Christakoglou, P., Christensen, C., Christiansen, P., Chujo, T., Chung, S., Cicalo, C., Cifarelli, L., Cindolo, F., Claus, G., Cleymans, J., Colamaria, F., Colella, D., Coli, S., Colledani, C., Collu, A., Colocci, M., Conesa Balbastre, G., Valle, Z., Connors, M., Contin, G., Contreras, J., Cormier, T., Corrales Morales, Y., Cortese, P., Cortés Maldonado, I., Cosentino, M., Costa, F., Crochet, P., Cruz Albino, R., Cuautle, E., Cunqueiro, L., Dainese, A., Dang, R., Danu, A., Da Riva, E., Das, D., Das, I., Das, K., Das, S., Dash, A., Dash, S., De, S., Decosse, C., Delagrange, H., Deloff, A., Dénes, E., D'Erasmo, G., Barros, G., De Caro, A., Cataldo, G., Cuveland, J., De Falco, A., De Gruttola, D., De Marco, N., De Pasquale, S., De Robertis, G., De Roo, K., Rooij, R., Diaz Corchero, M., Dietel, T., Divià, R., Di Bari, D., Di Liberto, S., Di Mauro, A., Di Nezza, P., Djuvsland, O., Dobrin, A., Dobrowolski, T., Domenicis Gimenez, D., Dönigus, B., Dordic, O., Dorheim, S., Dorokhov, A., Doziere, G., Dubey, A., Dubla, A., Ducroux, L., Dulinski, W., Dupieux, P., Dutta Majumdar, A., Ehlers III, R., Elia, D., Engel, H., Erazmus, B., Erdal, H., Eschweiler, D., Espagnon, B., Estienne, M., Esumi, S., Evans, D., Evdokimov, S., Eyyubova, G., Fabris, D., Faivre, J., Falchieri, D., Fantoni, A., Fasel, M., Fehlker, D., Feldkamp, L., Felea, D., Feliciello, A., Feofilov, G., Ferencei, J., Fernández Téllez, A., Ferreiro, E., Ferretti, A., Festanti, A., Figiel, J., Figueredo, M., Filchagin, S., Finogeev, D., Fionda, F., Fiore, E., Fiorenza, G., Floratos, E., Floris, M., Foertsch, S., Foka, P., Fokin, S., Fragiacomo, E., Francescon, A., Franco, M., Frankenfeld, U., Fuchs, U., Furget, C., Fusco Girard, M., Gaardhoje, J., Gagliardi, M., Gajanana, D., Gallio, M., Gangadharan, D., Ganoti, P., Garabatos, C., Garcia-Solis, E., Gargiulo, C., Garishvili, I., Gerhard, J., Germain, M., Gheata, A., Gheata, M., Ghidini, B., Ghosh, P., Ghosh, S., Gianotti, P., Giubilato, P., Giubellino, P., Gladysz-Dziadus, E., Glässel, P., Gomez, R., Gomez Marzoa, M., González-Zamora, P., Gorbunov, S., Görlich, L., Gotovac, S., Graczykowski, L., Grajcarek, R., Greiner, L., Grelli, A., Grigoras, A., Grigoras, C., Grigoriev, V., Grigoryan, A., Grigoryan, S., Grinyov, B., Grion, N., Grondin, D., Grosse-Oetringhaus, J., Grossiord, J., Grosso, R., Guber, F., Guernane, R., Guerzoni, B., Guilbaud, M., Gulbrandsen, K., Gulkanyan, H., Gunji, T., Gupta, A., Gupta, R., H Khan, K., Haake, R., Haaland, O., Hadjidakis, C., Haiduc, M., Hamagaki, H., Hamar, G., Hanratty, L., Hansen, A., Harris, J., Hartmann, H., Harton, A., Hatzifotiadou, D., Hayashi, S., Heckel, S., Heide, M., Helstrup, H., Hennes, E., Herghelegiu, A., Herrera Corral, G., Hess, B., Hetland, K., Hicks, B., Hillemanns, H., Himmi, A., Hippolyte, B., Hladky, J., Hristov, P., Huang, M., Hu-Guo, C., Humanic, T., Hutter, D., Hwang, D., Igolkin, S., Ijzermans, P., Ilkaev, R., Ilkiv, I., Inaba, M., Incani, E., Innocenti, G., Ionita, C., Ippolitov, M., Irfan, M., Ivanov, M., Ivanov, V., Ivanytskyi, O., Jacholkowski, A., Jadlovsky, J., Jahnke, C., Jang, H., Janik, M., Jayarathna, P., Jena, S., Jimenez Bustamante, R., Jones, P., Jung, H., Junique, A., Jusko, A., Kalcher, S., Kalinak, P., Kalweit, A., Kamin, J., Kang, J., Kaplin, V., Kar, S., Karasu Uysal, A., Karavichev, O., Karavicheva, T., Karpechev, E., Kebschull, U., Keidel, R., Keil, M., Ketzer, B., Khan, M., Khan, P., Khan, S., Khanzadeev, A., Kharlov, Y., Kileng, B., Kim, B., Kim, D., Kim, D., Kim, D., Kim, J., Kim, M., Kim, M., Kim, S., Kim, T., Kirsch, S., Kisel, I., Kiselev, S., Kisiel, A., Kiss, G., Klay, J., Klein, J., Klein-Bösing, C., Kluge, A., Knichel, M., Knospe, A., Kobdaj, C., Kofarago, M., Köhler, M., Kollegger, T., Kolojvari, A., Kondratiev, V., Kondratyeva, N., Konevskikh, A., Kovalenko, V., Kowalski, M., Kox, S., Koyithatta Meethaleveedu, G., Kral, J., Králik, I., Kramer, F., Kravcáková, A., Krelina, M., Kretz, M., Krivda, M., Krizek, F., Krus, M., Krymov, E., Kryshen, E., Krzewicki, M., Kucera, V., Kucheriaev, Y., Kugathasan, T., Kuhn, C., Kuijer, P., Kulakov, I., Kumar, J., Kurashvili, P., Kurepin, A., Kurepin, A., Kuryakin, A., Kushpil, S., Kushpil, V., Kweon, M., Kwon, Y., Guevara, P., Lagana Fernandes, C., Lakomov, I., Langoy, R., Lara, C., Lardeux, A., Lattuca, A., La Pointe, S., La Rocca, P., Lea, R., Lee, G., Legrand, I., Lehnert, J., Lemmon, R., Lenhardt, M., Lenti, V., Leogrande, E., Leoncino, M., León Monzón, I., Lesenechal, Y., Lévai, P., Li, S., Lien, J., Lietava, R., Lindal, S., Lindenstruth, V., Lippmann, C., Lisa, M., Listratenko, O., Ljunggren, H., Lodato, D., Loddo, F., Loenne, P., Loggins, V., Loginov, V., Lohner, D., Loizides, C., Lopez, X., López Torres, E., Lu, X., Luettig, P., Lunardon, M., Luo, J., Luparello, G., Luzzi, C., M Gago, A., M Jacobs, P., Ma, R., Maevskaya, A., Mager, M., Mahapatra, D., Maire, A., Malaev, M., Maldonado Cervantes, I., Malinina, L., Mal'Kevich, D., Maltsev, N., Malzacher, P., Mamonov, A., Manceau, L., Manko, V., Manso, F., Manzari, V., Mapelli, A., Marchisone, M., Mares, J., Margagliotti, G., Margotti, A., Marín, A., Marin Tobon, C., Markert, C., Marquard, M., Marras, D., Martashvili, I., Martin, N., Martinengo, P., Martínez, M., Martínez García, G., Martin Blanco, J., Martynov, Y., Mas, A., Masciocchi, S., Masera, M., Maslov, M., Masoni, A., Massacrier, L., Mastroserio, A., Mattiazzo, S., Matyja, A., Mayer, C., Mazer, J., Mazumder, R., Mazza, G., Mazzoni, M., Meddi, F., Menchaca-Rocha, A., Mercado Pérez, J., Meres, M., Miake, Y., Mikhaylov, K., Milano, L., Milosevic, J., Mischke, A., Mishra, A., Miskowiec, D., Mitu, C., Mlynarz, J., Mohanty, B., Molnar, L., Mongelli, M., Montaño Zetina, L., Montes, E., Morando, M., Moreira De Godoy, D., Morel, F., Moretto, S., Morreale, A., Morsch, A., Muccifora, V., Mudnic, E., Muhammad Bhopal, F., Muhuri, S., Mukherjee, M., Müller, H., Munhoz, M., Murray, S., Musa, L., Musinsky, J., Nandi, B., Nania, R., Nappi, E., Nattrass, C., Nayak, T., Nazarenko, S., Nedosekin, A., Nicassio, M., Niculescu, M., Nielsen, B., Nikolaev, S., Nikulin, S., Nikulin, V., Nilsen, B., Noferini, F., Nomokonov, P., Nooren, G., Nyanin, A., Nystrand, J., Oeschler, H., Oh, S., Oh, S., Okatan, A., Olah, L., Oleniacz, J., Oliveira Da Silva, A., Onderwaater, J., Oppedisano, C., Ortiz Velasquez, A., Oskarsson, A., Otwinowski, J., Oyama, K., Pachmayer, Y., Pachr, M., Pagano, P., Paic, G., Painke, F., Pajares, C., Pal, S., Palmeri, A., Panati, S., Pant, D., Pantano, D., Papikyan, V., Pappalardo, G., Park, W., Passfeld, A., Pastore, C., Patalakha, D., Paticchio, V., Paul, B., Pawlak, T., Peitzmann, T., Pereira Da Costa, H., Pereira De Oliveira Filho, E., Peresunko, D., Pérez Lara, C., Peryt, W., Pesci, A., Pestov, Y., Petagna, P., Petrácek, V., Petran, M., Petris, M., Petrovici, M., Petta, C., Pham, H., Piano, S., Pikna, M., Pillot, P., Pinazza, O., Pinsky, L., Piyarathna, D., Ploskon, M., Planinic, M., Pluta, J., Pochybova, S., Podesta-Lerma, P., Poghosyan, M., Pohjoisaho, E., Polichtchouk, B., Poljak, N., Pop, A., Porteboeuf-Houssais, S., Porter, J., Pospisil, V., Potukuchi, B., Prasad, S., Preghenella, R., Prino, F., Protsenko, M., Pruneau, C., Pshenichnov, I., Puddu, G., Puggioni, C., Punin, V., Putschke, J., Qvigstad, H., Rachevski, A., Raha, S., Rak, J., Rakotozafindrabe, A., Ramello, L., Raniwala, R., Raniwala, S., Räsänen, S., Rascanu, B., Rasson, J., Rathee, D., Rauf, A., Razazi, V., Read, K., Real, J., Redlich, K., Reed, R., Rehman, A., Reichelt, P., Reicher, M., Reidt, F., Renfordt, R., Reolon, A., Reshetin, A., Rettig, F., Revol, J., Reygers, K., Riabov, V., Ricci, R., Richert, T., Richter, M., Riedler, P., Riegler, W., Riggi, F., Rivetti, A., Rocco, E., Rodríguez Cahuantzi, M., Rodriguez Manso, A., Roed, K., Rogochaya, E., Rohni, S., Rohr, D., Röhrich, D., Romita, R., Ronchetti, F., Ronflette, L., Rosnet, P., Rossegger, S., Rossewij, M., Rossi, A., Roudier, S., Rousset, J., Roy, A., Roy, C., Roy, P., Rubio Montero, A., Rui, R., Russo, R., Ryabinkin, E., Ryabov, Y., Rybicki, A., Sacchetti, M., Sadovsky, S., Safarík, K., Sahlmuller, B., Sahoo, R., Sahu, P., Saini, J., Salgado, C., Salzwedel, J., Sambyal, S., Samsonov, V., Sanchez Castro, X., Sánchez Rodríguez, F., Sándor, L., Sandoval, A., Sano, M., Santagati, G., Santoro, R., Sarkar, D., Scapparone, E., Scarlassara, F., Scharenberg, R., Schiaua, C., Schicker, R., Schipper, J., Schmidt, C., Schmidt, H., Schuchmann, S., Schukraft, J., Schulc, M., Schuster, T., Schutz, Y., Schwarz, K., Schweda, K., Scioli, G., Scomparin, E., Scott, P., Scott, R., Segato, G., Seger, J., Selyuzhenkov, I., Senyukhov, S., Seo, J., Serradilla, E., Sevcenco, A., Sgura, I., Shabetai, A., Shabratova, G., Shahoyan, R., Shangaraev, A., Sharma, N., Sharma, S., Shigaki, K., Shtejer, K., Sibiriak, Y., Siddhanta, S., Siemiarczuk, T., Silvermyr, D., Silvestre, C., Simatovic, G., Singaraju, R., Singh, R., Singha, S., Singhal, V., Sinha, B., Sinha, T., Sitar, B., Sitta, M., Skaali, T., Skjerdal, K., Smakal, R., Smirnov, N., Snellings, R., Snoeys, W., Sogaard, C., Soltz, R., Song, J., Song, M., Sooden, V., Soramel, F., Sorensen, S., Spacek, M., Spalek, J., Spiriti, E., Sputowska, I., Spyropoulou-Stassinaki, M., Srivastava, B., Stachel, J., Stan, I., Stefanek, G., Steinpreis, M., Stenlund, E., Steyn, G., Stiller, J., Stocco, D., Stolpovskiy, M., Strmen, P., Suaide, A., Subieta Vasquez, M., Sugitate, T., Suire, C., Suleymanov, M., Suljic, M., Sultanov, R., Sumbera, M., Sun, X., Susa, T., Symons, T., Toledo, A., Szarka, I., Szczepankiewicz, A., Szymanski, M., Takahashi, J., Tangaro, M., Tapia Takaki, J., Tarantola Peloni, A., Tarazona Martinez, A., Tauro, A., Tejeda Muñoz, G., Telesca, A., Terrevoli, C., Ter Minasyan, A., Thäder, J., Thomas, D., Tieulent, R., Timmins, A., Toia, A., Torii, H., Trubnikov, V., Trzaska, W., Tsuji, T., Tumkin, A., Turchetta, R., Turrisi, R., Tveter, T., Tymchuk, I., Ulery, J., Ullaland, K., Uras, A., Usai, G., Vajzer, M., Vala, M., Valencia Palomo, L., Valentino, V., Valin, I., Vallero, S., Vande Vyvre, P., Vannucci, L., Van Der Maarel, J., Van Hoorne, J., Leeuwen, M., Vargas, A., Varma, R., Vasileiou, M., Vasiliev, A., Vasta, P., Vechernin, V., Veldhoen, M., Velure, A., Venaruzzo, M., Vercellin, E., Vergara Limón, S., Verlaat, B., Vernet, R., Verweij, M., Vickovic, L., Viesti, G., Viinikainen, J., Vilakazi, Z., Villalobos Baillie, O., Vinogradov, A., Vinogradov, L., Vinogradov, Y., Virgili, T., Viyogi, Y., Vodopyanov, A., Völkl, M., Voloshin, K., Voloshin, S., Volpe, G., Haller, B., Vorobyev, I., Vranic, D., Vrláková, J., Vulpescu, B., Vyushin, A., Wagner, B., Wagner, J., Wagner, V., Wang, M., Wang, Y., Watanabe, D., Weber, M., Wessels, J., Westerhoff, U., Wiechula, J., Wikne, J., Wilde, M., Wilk, G., Wilkinson, J., Williams, M., Windelband, B., Winn, M., Winter, M., Xiang, C., Yaldo, C., Yamaguchi, Y., Yang, H., Yang, P., Yang, S., Yano, S., Yasnopolskiy, S., Yi, J., Yin, Z., Yoo, I., Yushmanov, I., Zaccolo, V., Zach, C., Zaman, A., Zampolli, C., Zaporozhets, S., Zarochentsev, A., Závada, P., Zaviyalov, N., Zbroszczyk, H., Zgura, I., Zhalov, M., Zhang, F., Zhang, H., Zhang, X., Zhang, Y., Zhao, C., Zherebchevsky, V., Zhou, D., Zhou, F., Zhou, Y., Zhu, H., Zhu, J., Zhu, X., Zichichi, A., Zimmermann, A., Zimmermann, M., Zinovjev, G., Zoccarato, Y., Zynovyev, M. \& Zyzak, M. Technical Design Report for the Upgrade of the ALICE Inner Tracking System.  (2014), https://cds.cern.ch/record/1625842
\bibitem{ITS3}The, A. Technical Design report for the ALICE Inner Tracking System 3 - ITS3 ; A bent wafer-scale monolithic pixel detector. (CERN,2024), https://cds.cern.ch/record/2890181, Co-project Manager: Magnus Mager, magnus.mager@cern.chds



\end{thebibliography}
\end{document}